# Effect of Orientation of Cation CH$_3$NH$_3$PbI$_3$ on Ambipolarity, Open-Circuit Voltage, and Excitons' Lifetime

K. Ouassoul[1,2], A. El Kenz[2], M. Loulidi[2], A. Benyoussef[3], M. Azzouz[1,4,*]

[1]*International University of Rabat, LERMA, School of Aerospace Engineering, Sala El Jadida, Morocco*

[2]*Laboratory of Condensed Matter and Interdisciplinary Sciences, Mohammed V-Agdal. University, Rabat, Morocco*

[3]*Hassan II Academy of Sciences and Techniques, Rabat, Morocco*

[4]*Department of Physics, Laurentian University, Ramsey Lake Road, Sudbury, Ontario P3E 2C6, Canada*

**\*Corresponding author: mohamed.azzouz@uir.ac.ma; mazzouz@laurentian.ca**

## Abstract

The effect of the orientation of the cation MA (i.e., CH$_3$NH$_3$) on the physical properties of MAPbI$_3$ are investigated using the density functional theory. We report that the Fermi energy level exhibits a large variation with MA orientation, and it is this Fermi level variation that makes this material unique and so different from the non-organic perovskites. The Fermi level variation with orientation of MA is proposed to be responsible for the experimentally observed intrinsic open-circuit voltage $V_{OC}$, ambipolarity, and long lifetime of the excitons. Based on our results, ferroelectric domains in the low-temperature orthorhombic phase or clusters of MA molecules not rotating in unison (as a consequence of e.g. impurities and/or defaults) in the higher-temperature tetragonal and cubic phases are proposed to be responsible for ambipolarity. This is because any given two adjacent domains or clusters with different MA's average orientations may have very different Fermi levels, causing them to behave as an effective "diode". Also, the measured significant lifetime of excitons in MAPbI$_3$ finds a natural explanation using this effective diode. Thus, our work provides a unified manner for explaining at least these three important properties in terms of the rotation-induced Fermi level variations. We used the density functional theory to get our results.

## I.      Introduction

Organic-inorganic hybrid perovskites with prototype formula MAPbI$_3$, where the cation MA stands for CH$_3$NH$_3$, have been intensively studied since 2009 for solar cells' applications [1]. The power conversion efficiency (PCE) of these cells showed an extraordinary increase from 3.81% in the first use [1] to more than 25.2 % recently [2], while the efficiency of inorganic perovskites is ~11% actually [3,4]. Furthermore, the organic perovskite materials are easier to synthesize from cheaper raw materials with low production costs [5]. MAPbI$_3$ shows interesting and attractive characteristics, important optoelectronic properties such as a band gap in the visible spectrum, and small effective masses for both electrons and holes [6]. This material is also characterized by large diffusion lengths, namely ~130 nm for electrons and ~90 nm for holes [7], and by long carrier lifetime of ~0.5 µs [8]. In the present work, we analyzed three very important quantities, for which we summarized the most relevant literature as follows:

Excitons' lifetime: The intrinsic dipole of the cation MA is thought to generate ferroelectric domains that are responsible for the long-lifetime of photoexcited electron-hole pairs; i.e., the slow recombination of holes and electrons is due to the ferroelectric domains [9]. Note that ferroelectricity in the present hybrid perovskite is an issue that several reports investigated at room temperature. Some experimental reports refuted ferroelectricity in this material [10,11], while another one supported antiferroelectricity [12]. Other reports argued for the occurrence of ferroelasticity [13] or pyroelectricity [14].

A recent study related the long lifetime to multi-energy band gaps in the structure leading to the absorption of more photons with different energies, and the variation of the gap nature reduces significantly the recombination of electrons and holes [15]. It has also been suggested that a large electrostatic potential fluctuation due to the random orientation of MA cations localizes the conduction band minimum (CBM) in the low electrostatic potential region and the valence band maximum (VBM) in the high electrostatic region. This separates the charges and thus increases carriers' lifetime [16].

It is also thought that electron-hole recombination is reduced because the band edges shift apart from each other in the Brillouin zone due to Rashba effect [17]. Despite this, the carrier mobility is modest (67.2 cm$^2$ V$^{-1}$ s$^{-1}$ according to Saidaminov *et al.* [8] and 66 cm$^2$ V$^{-1}$ s$^{-1}$ by Stoumpos *et al.* [18]) when compared to inorganic semiconductors [19]. It has been suggested that the electron mobility is reduced by a factor two because of large polarons [20]. The carrier transport is also thought to be affected by the cations' rotational disorder and the soft inorganic cage through electron-phonon coupling [21]. This material has significant light absorption in the visible to near-infrared (IR) range [6]. However, the binding energy of photon-generated excitons in MAPbI$_3$ is low. Experimentally, Grätzel determined the binding energy to be about 30 meV [22], and a theoretical study by Hakamata *et al.* showed that the time averaged exciton's binding energy is $12 \pm 9$ meV [23].

Ambipolarity: Canicoba *et al.* proposed that the alteration from a p-type field effect transistor (FET) to an ambipolar FET is due to the ions local redistribution after a continuous gate voltage cycling, which creates domains of n-type and p-type in the channel of the perovskite FET [24]. In this regard, Weil Yu *et al.* observed also the FET ambipolarity behavior, and they attributed it to the fabrication method without giving any physical explanation of the transformation mechanism [25]. Theoretically, Giorgi *et al.* [26] reported that the cation MA is responsible for the ambipolarity in the cubic phase without mentioning any fundamental mechanism related to the Fermi energy variations due to MA rotation. The ambipolarity was confirmed by Li *et al.* [27] using MAPbI$_3$ in phototransistor applications. It was also reported that the charge transport can be either p or n type depending on the synthesis methods used, which are characterized by different levels of defects [18]. In this work, we attribute the ambipolarity to the rotation of the MA molecule and the large Fermi level variation it induces.

Open-circuit voltage: Patel *et al* published the variation of the open-circuit voltage $V_{OC}$ with temperature [28]. They found that $V_{OC}$ varies strongly in the orthorhombic phase below 162 K. $V_{OC}$ changes however only little between 0.7 V and 1.13 V in the tetragonal and cubic phase over a very large temperature range. $V_{OC}$ becomes much smaller in the ferroelectric domain at very low temperature, where the MA molecules tend to form a single ferroelectric domain in which they are all aligned in the same direction on average. This seems to suggest that the free rotation of MA molecules or the presence of more than one domain in the tetragonal and cubic phases is responsible for the larger values of $V_{OC}$. This idea is examined further in this paper.

The rotation of MA cations in the inorganic cage of MAPbI$_3$ turns out to be responsible for the above three properties. The material MAPbI$_3$ adopts the cubic structure above temperature

327.4 K with space group Pm3m [29]. In this phase, Nuclear Magnetic Resonance (NMR) [30] showed that MA cations rotate with picosecond scale dynamics and are dynamically disordered, moving in an isotropic potential at a rate approaching that of the freely rotating MA cation [29]. Also, Molecular dynamics simulations show, using a reasonably large supercell, that the correlation between MA ions is low at room temperature [31]. Moreover, the precise role of organic cations in the high-power conversion efficiency is still a debated issue [32]–[34].

In this work, we analyzed the effects of the rotation of MA cations on the three physical properties of MAPbI$_3$ discussed above. Our goals were to find a physical phenomenon (mechanism) able to explain simultaneously the long excitons' lifetime, open-circuit potential, and ambipolarity. We found that the rotation of MA cations in the cubic (so also in the tetragonal) phase is responsible for the long lifetime of excitons because the rotation of MA cations induces significant variation in the Fermi energy level. We conclude that in the orthorhombic phase, the ferroelectric domains should develop the same effect, i.e., any two adjacent domains with different orientations for the MA electric moment act as a "diode" with two different Fermi energies. This difference in the Fermi levels separate electrons and holes created by photons in the same way as in PN junctions. We also found that the difference between the maximum and minimum of the orientation-dependent Fermi energy creates a potential difference very close to the experimentally measured open-circuit voltage. Finally, the occurrence of ferroelectric domains is proposed to cause this material to be ambipolar because different domains with different MA orientations have different Fermi levels. They may thus act as semiconductors of different types.

This paper is organized as follows: In Section II we describe the computation methodology. Section III is devoted to results and discussions. We investigate the rotation-induced changes in the crystal structure, electronic structure, open-circuit voltage, ambipolarity, and exciton's lifetime. Then we calculate the optical conductivity. Conclusions are drawn in Section IV.

## II. Material and Methodology

In the framework of the density functional theory (DFT), we investigated the cubic phase of the MAPbI$_3$ system, which belongs in Pm3m space group, and contains 12 atoms in the primitive cell. We performed our calculations using the Quantum Espresso software [35]. The Projector Augmented Wave (PAW) pseudopotential was used for describing electron-ion interactions [36], and the Generalized Gradient Approximation (GGA) parameterized by Perdew-Burke-Ernzerhof (PBE) was used for treating the electron-electron exchange correlation functional. The Monkhorst Pack **k**-mesh grid of $8 \times 8 \times 8$ is used for all sample sizes of our calculations [37]. The Van Der Waals interaction was also considered because it yields lattice parameters comparable to experimental values [38]–[41]. To represent the wave functions, we used 60 Ry for the energy cut-off. The relaxation is satisfied when the atomic forces are less than 0.01 eV/Å. Further details are included in the supplemental material.

The transport properties were calculated using the Boltztrap code [42], which provides the transport response functions based on the resolution of Boltzmann's transport equation within the constant relaxation time approximation using Fourier expansion [42]. This equation plays an important role in the theoretical understanding of transport phenomena, i.e. the response of a system maintained in external conditions of disequilibrium (gradient of potential, temperature, concentration, or velocity).

### III. RESULTS AND DISCUSSION
#### A. Lattice parameters and cell angles versus MA's orientation

As already mentioned in the introduction, the dynamic behavior of the organic cation MA has been the subject of numerous studies. In the present work, we investigated the effect of MA rotations on the electronic structure; specifically, we addressed the role played by these rotations in the lifetime of the electron-hole (exciton) pairs, and in the open-circuit voltage that characterizes the solar cells based on MAPbI$_3$. The atomic unit cell is displayed in Fig. 1. For the cubic phase, the organic molecules MA are dynamically disordered, rotating in an isotropic potential at a rate approaching that of the freely rotating MA cation [29]. The MA cations' rotation around their C-N axes has a negligible effect; it only changes slightly the energy of the system (by ~2 meV) [43] and causes an insignificant variation of the lattice parameters [34]. Here, we rather focus on MA cations' rotations around their centers. We select six specific orientations in the plane (010) which are relative to axis **a**: angles 0°, 30°, 45°, 60°, 90°, and 180°, five orientations in the plane (100) relative to the **b** axis, namely angles 0°, 30°, 45°, 60°, 90°, and one orientation along direction [111]. Fig. 2 shows the different orientations of the cation MA in the **ac** plane relative to axis **a**.

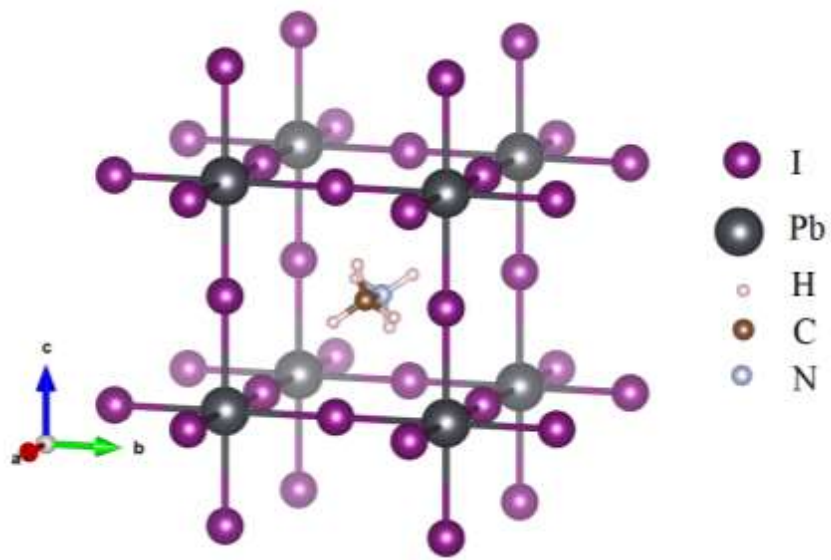

FIG. 1. The atomic unit cell of CH3NH3PbI3.

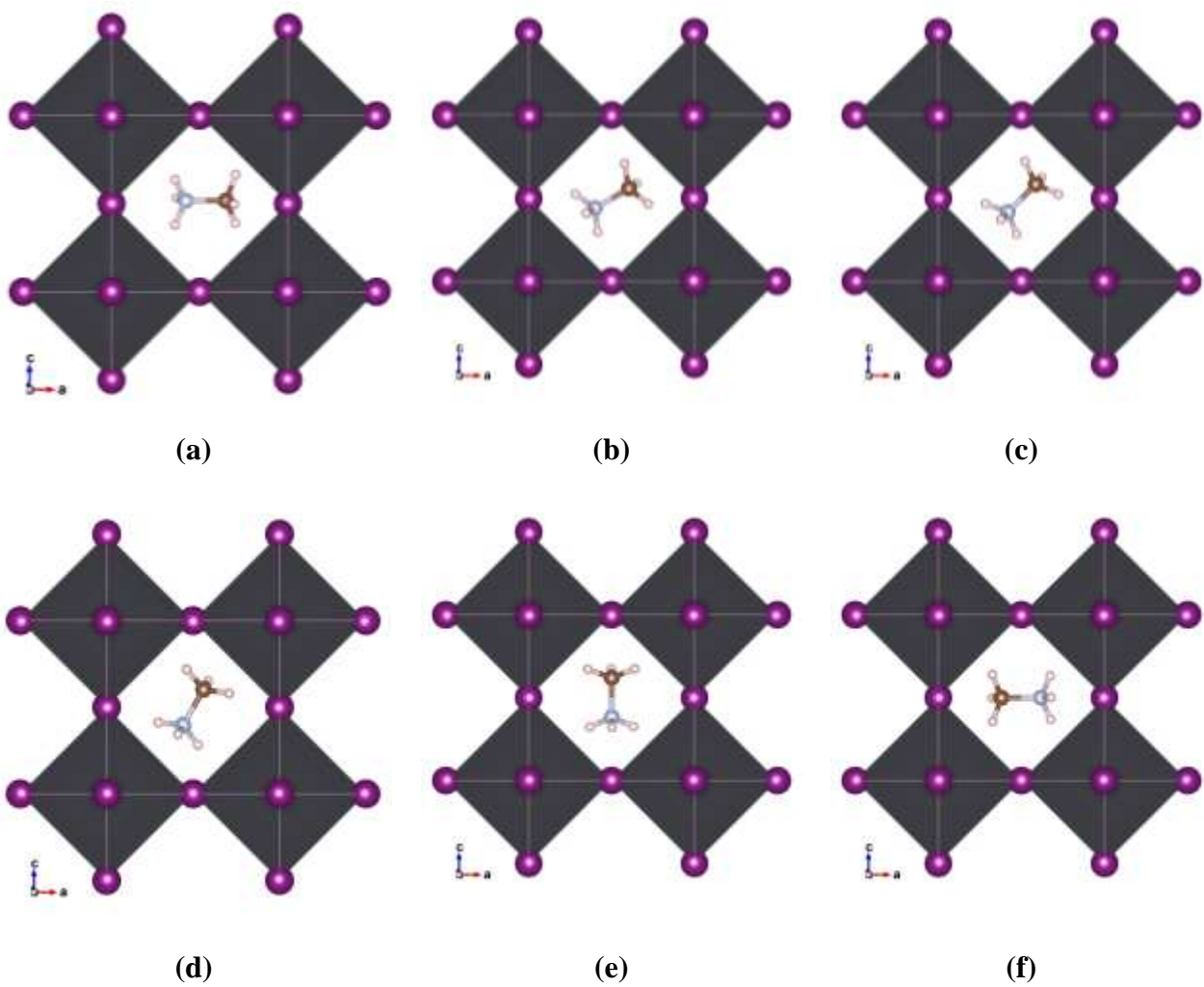

(a)　　　　　　　　　(b)　　　　　　　　　(c)

(d)　　　　　　　　　(e)　　　　　　　　　(f)

FIG. 2. The structures of MAPbI3 for different orientations of the cation MA relative to axis **a** in **ac** plane: a**)** angle 0°; b**)** angle 30°; c) angle 45°; d) angle 60°; e) angle 90°, and f) angle 180°.

TABLE I. Cell parameters and angles after relaxation. (wrt means with respect to)

| Orientation | $a$ (Å) | $b$ (Å) | $c$ (Å) | $\alpha$ (°) | $\beta$ (°) | $\gamma$ (°) | Volume (Å$^3$) |
|---|---|---|---|---|---|---|---|
| 0° wrt [111] | 6.24 | 6.20 | 6.20 | 89.40 | 87.69 | 87.64 | 239.60 |
| In the plane (010) | | | | | | | |
| 0° wrt [100] | 6.24 | 6.20 | 6.21 | 90.03 | 90.62 | 90.05 | 240.32 |
| 30° wrt [100] | 6.23 | 6.17 | 6.22 | 88.60 | 87.13 | 90.63 | 238.49 |
| 45° wrt [100] | 6.22 | 6.16 | 6.22 | 89.85 | 87.37 | 89.75 | 238.04 |
| 60° wrt [100] | 6.21 | 6.17 | 6.23 | 90.51 | 87.88 | 88.38 | 238.17 |
| 90° wrt [100] | 6.19 | 6.18 | 6.27 | 90.01 | 89.33 | 89.98 | 239.93 |
| 180° wrt [100] | 6.24 | 6.22 | 6.20 | 90.04 | 89.94 | 89.33 | 240.38 |
| In the plane (100) | | | | | | | |
| 0° wrt [010] | 6.19 | 6.27 | 6.18 | 90.03 | 89.97 | 89.33 | 240.01 |
| 30° wrt [010] | 6.17 | 6.23 | 6.21 | 87.81 | 91.56 | 89.51 | 238.10 |
| 45° wrt [010] | 6.16 | 6.22 | 6.22 | 87.37 | 90.15 | 90.25 | 238.04 |
| 60° wrt [010] | 6.17 | 6.21 | 6.23 | 87.87 | 89.48 | 91.62 | 238.17 |
| 90° wrt [010] | 6.19 | 6.18 | 6.27 | 90.01 | 89.33 | 89.98 | 239.93 |
| XRD [18] | 6.3115 | 6.3115 | 6.3161 | 90.00 | 90.00 | 90.00 | 251.60 |

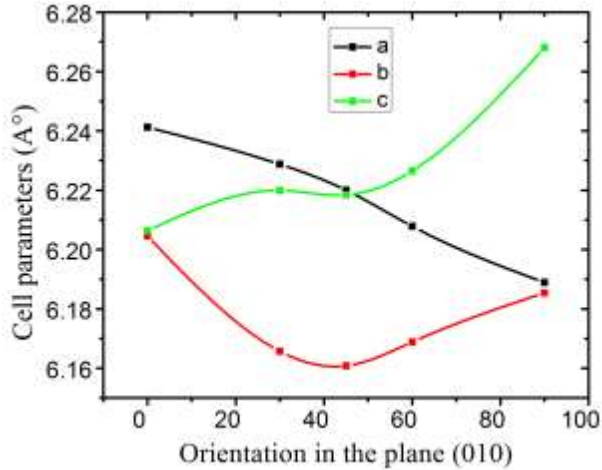
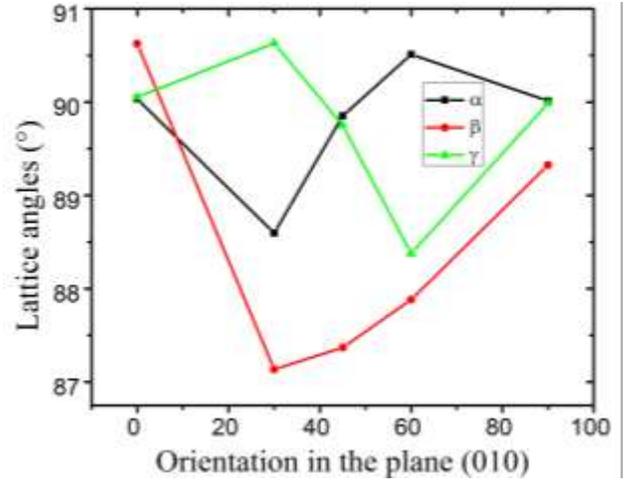

(a)

(b)

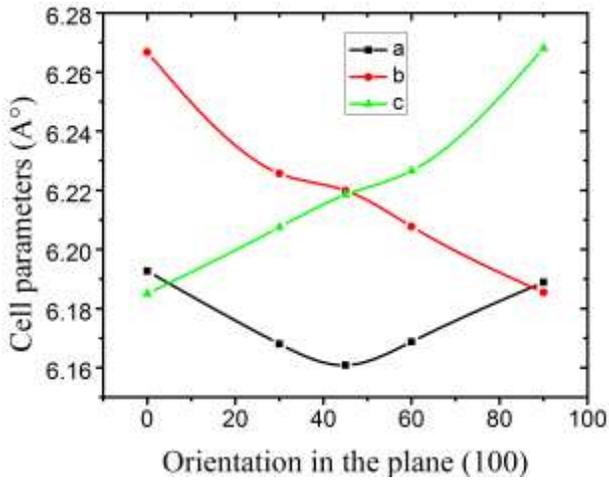
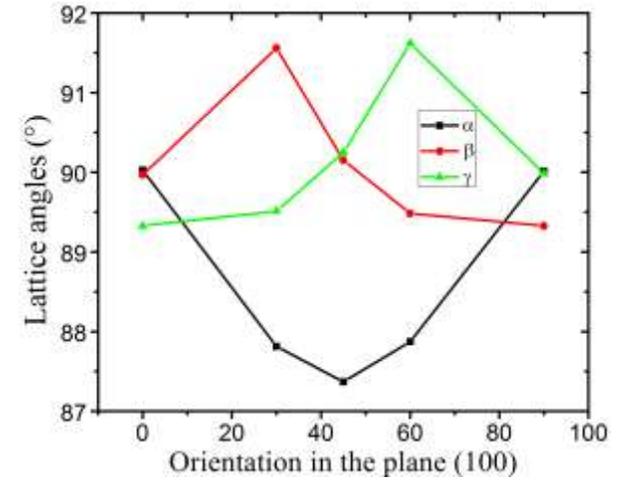

(c)

(d)

FIG. 3. Values of cell lattice parameters and angles of the cell as a function of orientation: (a) and (b) in the plane (010); (c) and (d) in the plane (100).

All structures with different orientations have been optimized using Broyden–Fletcher–Goldfarb–Shanno (BFGS) method [44]. According to the values of lattice parameters as a function of orientation, we found that in plane (010) the relaxed value of parameter $a$ is the largest when MA is oriented along axis **a**, and its value reduces gradually until orientation 90°. In the latter orientation, MA is oriented along **c** axis and the parameter $c$ becomes the largest relaxed value. The same behavior is observed in the plane (100) when MA's orientation varies from direction **b** to **c**. As shown in Fig. 3, the behavior of the curves for the orientations in the plane (010) is qualitatively the same as in the plane (100). MA pushes apart the atoms found along its axis. About

1.75% variation has been observed in the lattice parameters of the system and about 5.1% in lattice angles. The atoms Pb and I are in the faces of the cell; when the organic cation rotates, the positions of Pb and I change, thereby modifying slightly the lattice parameters. This affects the values of the lattice angles and unit cell volume as well. The crystal structure becomes pseudo-cubic. As seen in Table I, the maximum deviation between the calculated lattice parameters and the X-ray diffraction (XRD) experimental [18] value is 2.5%, which is reasonable for DFT calculations. Note that the values measured experimentally for the lattice parameters and cells angles are averages about all orientations in the high-temperature cubic phase of the compound $MAPbI_3$.

### B. Band structure

Fig. 4 displays the band structure for several orientations of MA, where energies are shifted by the valence band maximum (VBM). Our results indicate that without considering the effect of spin-orbit in our calculations, the bandgap remains direct at the R **k**-point for all orientations. The relaxation-induced modification of the symmetry, as we discussed earlier, changes only slightly the value of the bandgap, which remains close to the experimental value (1.5 eV) [45]. We also extracted the bandwidth dispersion in the valence band maximum for **k**-space directions Γ$X$ and Γ$M$ for all our MA orientations (Table S1). Our values compare well to experimental and other theoretical values [46].

The orientations along 45° with respect direction [010] in plane (100), and with respect direction [100] in plane (010) have the biggest values of the Fermi energy and the lowest bandgap energies among all orientations. For these two orientations, the lowest three bands in the conduction band get closer towards lower energies, and a reduction in the primitive cell volume and in the bandgap is observed as seen in Figs. 5(a) and 5(b). These results are consistent with those reported in Ref. [34] in which Mehdizadeh *et al.* explain that the rotation of MA affects the interaction between $PbI_3$ and MA. Any decrease in this interaction increases the bandgap and the cell volume as well.

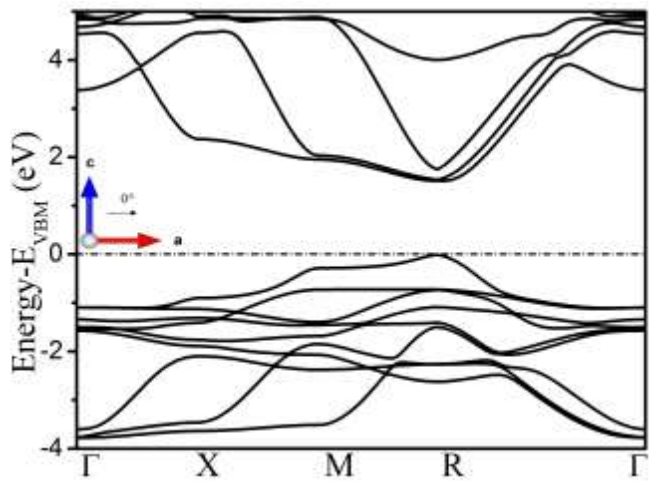

(a)

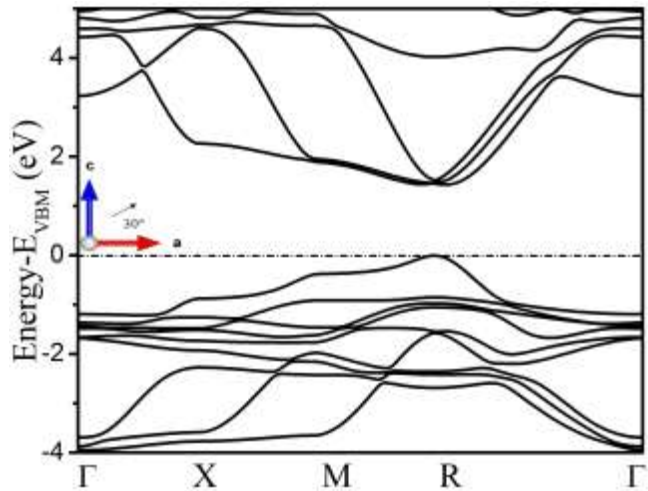

(b)

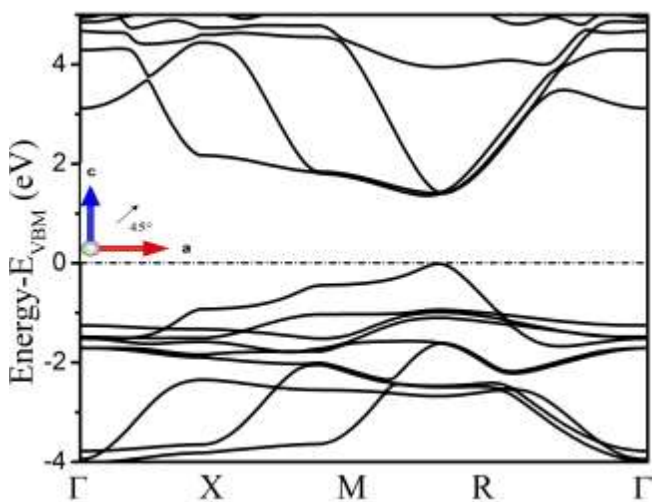

(c)

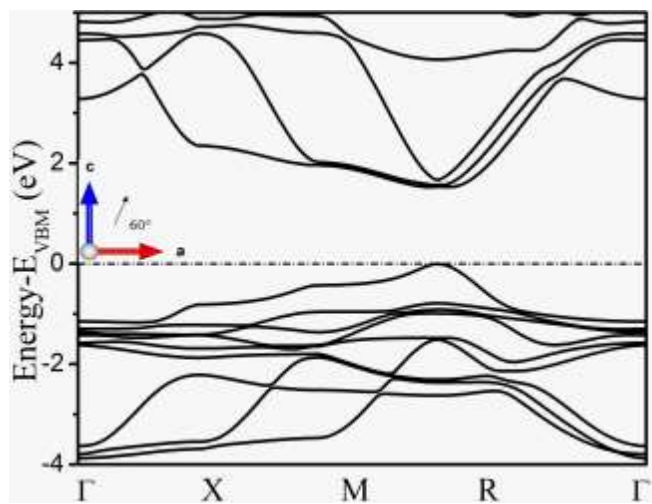

(d)

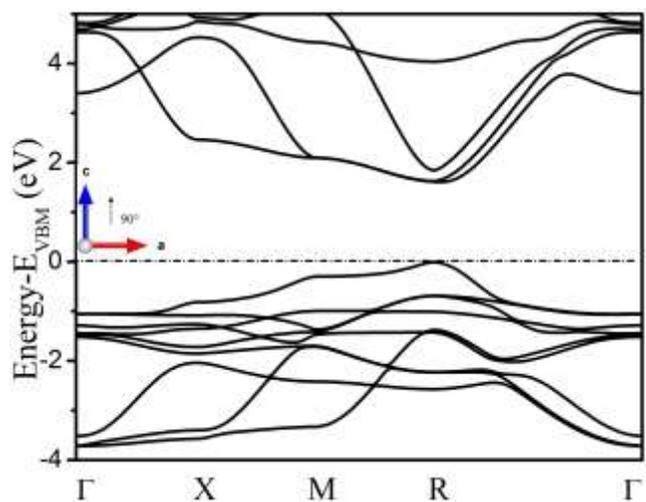

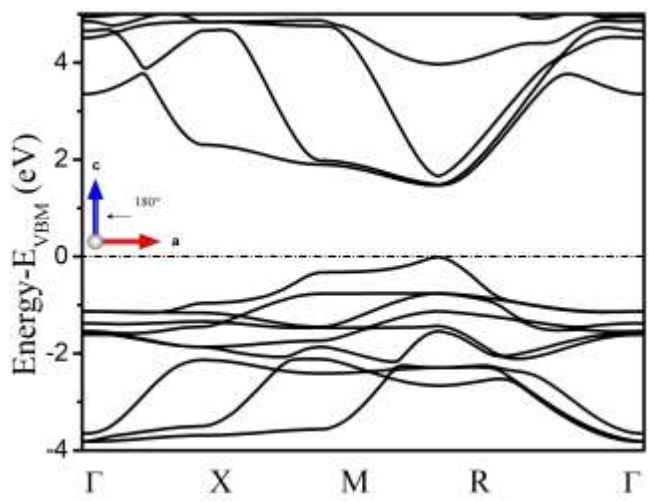

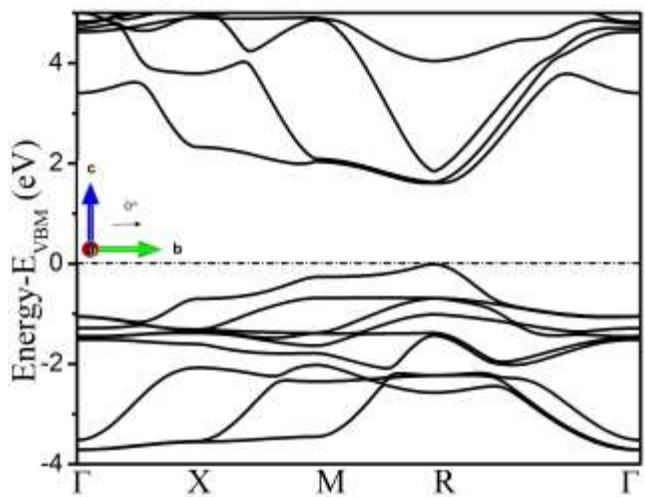

(e)

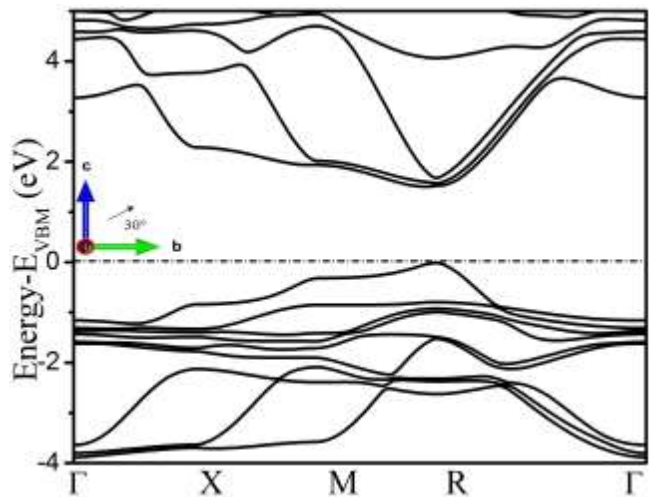

(f)

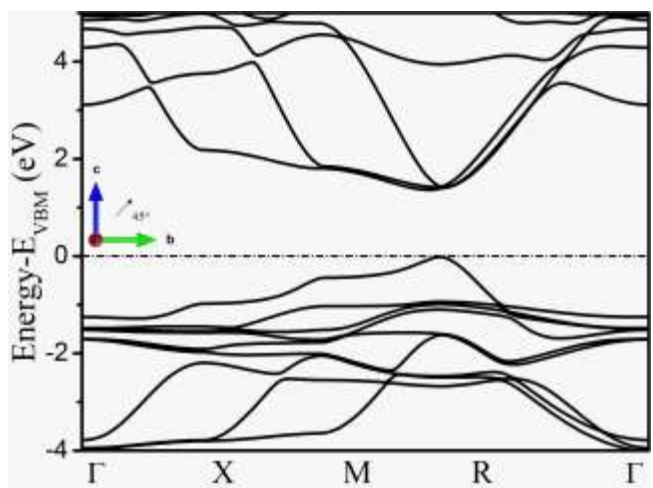

(g)

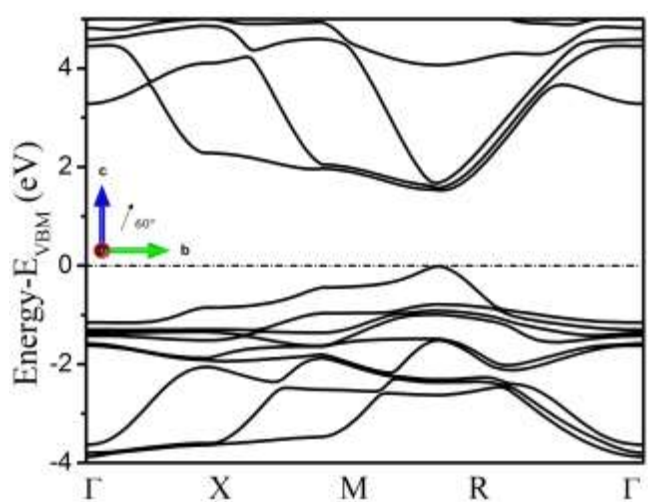

(h)

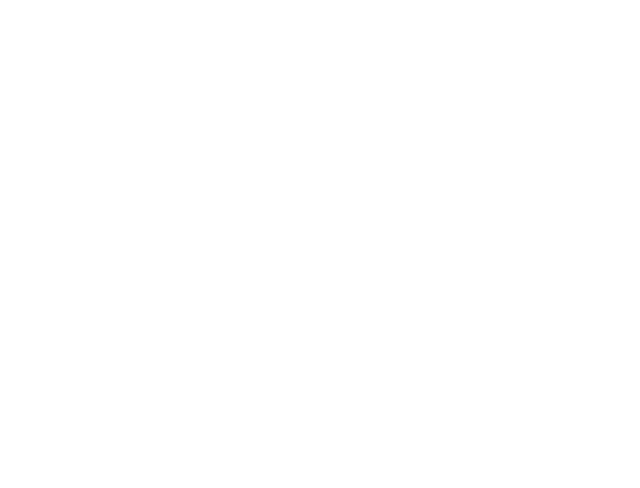

(i)

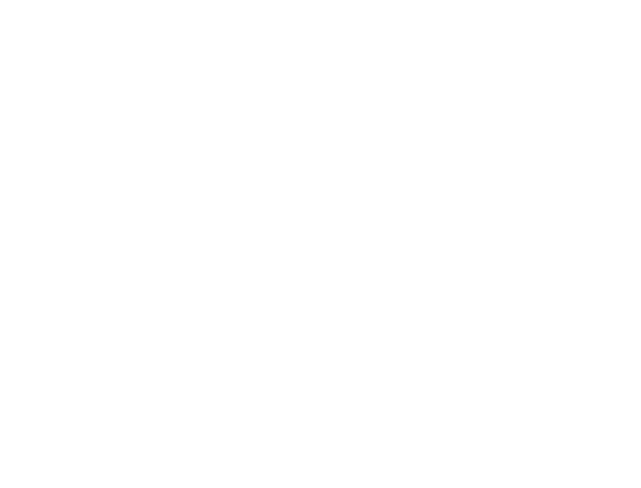

(j)

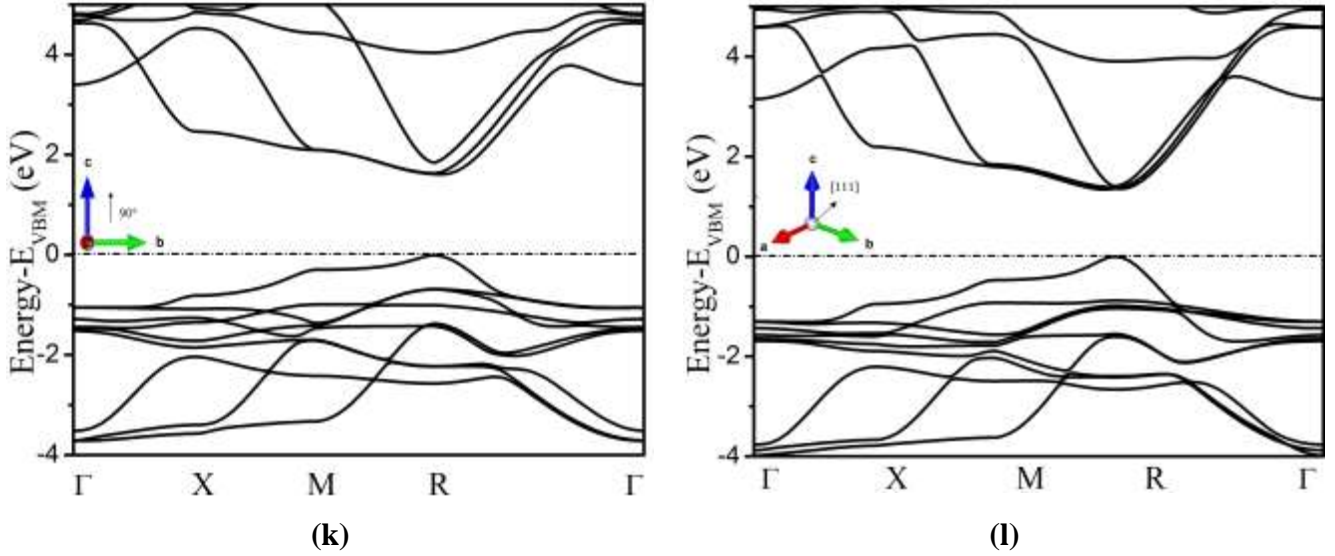

FIG. 4. The band structure of MAPbI3 for different orientations of the cation MA: a**)** angle 0°; b**)** angle 30°; c) angle 45°; d) angle 60°; e) angle 90°; f) angle 180° relative to axis **a** in the plane (010). g) angle 0°; h**)** angle 30°; i) angle 45°; j) angle 60°; k) orientation 90° relative to axis **b** in the plane (100). l) orientation [111]. The thin arrows in the insets refer to MA orientation and the arrowhead corresponds to carbon atom.

## C. Bandgap and cell volume versus MA's orientation

We calculated the bandgap and cell volume as a function of MA orientation. In Figs. 5(a) and 5(b), a direct relation between the bandgap and the cell volume as a function of orientations can be seen. This is in good agreement with literature, concerning the behavior and symmetry similarities about 45° [34]. The bandgap and cell volume minima occur at orientation 45° and show the same trends away from this orientation.

According to Schulz and co-workers [47], the valence band maximum (VBM) at 3.7 eV and the valence band minimum (CBM) at 5.4 eV are consistent with values for the orientation along **c** axis as shown in Table II. In this table the total energy, Fermi energy, band gap, energy barrier $\Delta E$, VBM, and CBM are listed for all MA orientations considered in this work. The energy barrier $\Delta E$ is the difference between the energy corresponding to the given orientation and the lowest energy which corresponds to orientation 30° with respect to [100] in plane (010).

TABLE II. Total energy, Fermi energy, band gap, energy barrier, VBM and CBM values for different orientations.

| Orientation | Energy (eV) | Fermi energy (eV) | Bandgap (eV) | ΔE (eV) | VBM (eV) | CBM (eV) |
|---|---|---|---|---|---|---|
| 0° wrt [111] | -34890.95888 | 4.225 | 1.3596 | 0.01974 | 3.9527 | 5.3123 |
| In the plane (010) | | | | | | |
| 0° wrt [100] | -34890.95347 | 4.1113 | 1.5232 | 0.02516 | 3.789 | 5.3122 |
| 30° wrt [100] | -34890.97862 | 4.1584 | 1.538 | 0 | 3.9352 | 5.4732 |
| 45° wrt [100] | -34890.96824 | 4.974 | 1.3923 | 0.01039 | 4.0593 | 5.4516 |
| 60° wrt [100] | -34890.97421 | 4.1567 | 1.5421 | 0.00441 | 3.9263 | 5.4684 |
| 90° wrt [100] | -34890.95071 | 4.1591 | 1.6164 | 0.02792 | 3.7576 | 5.374 |
| 180° wrt [100] | -34890.95324 | 4.1034 | 1.4661 | 0.02539 | 3.8223 | 5.2884 |
| In the plane (100) | | | | | | |
| 0° wrt [010] | -34890.95147 | 4.1567 | 1.6152 | 0.01039 | 3.7559 | 5.3711 |
| 30° wrt [010] | -34890.97355 | 4.1658 | 1.538 | 0.00507 | 3.9352 | 5.4732 |
| 45° wrt [010] | -34890.96826 | 4.9748 | 1.3924 | 0.01037 | 4.0591 | 5.4515 |
| 60° wrt [010] | -34890.9742 | 4.1567 | 1.5417 | 0.00441 | 3.9267 | 5.4684 |
| 90° wrt [010] | -34890.95071 | 4.1591 | 1.6164 | 0.02792 | 3.7576 | 5.374 |

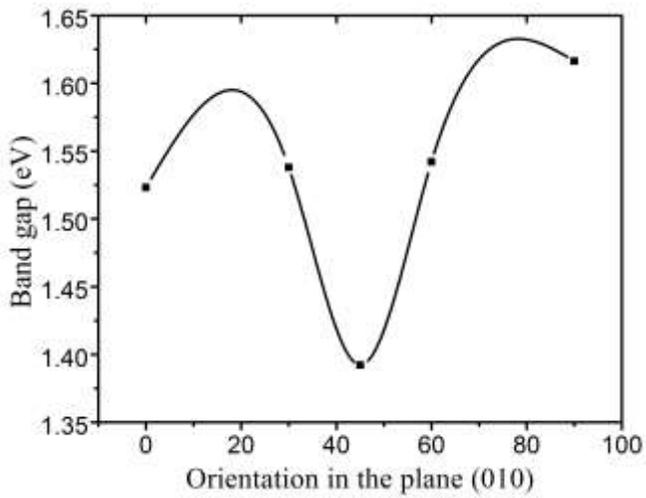
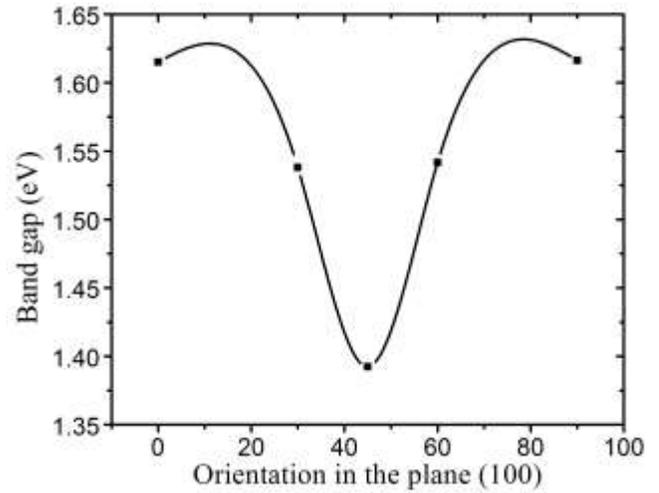

(a)

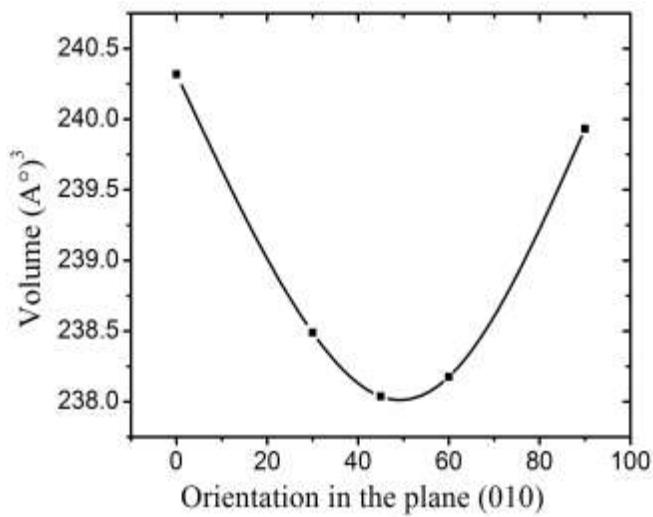
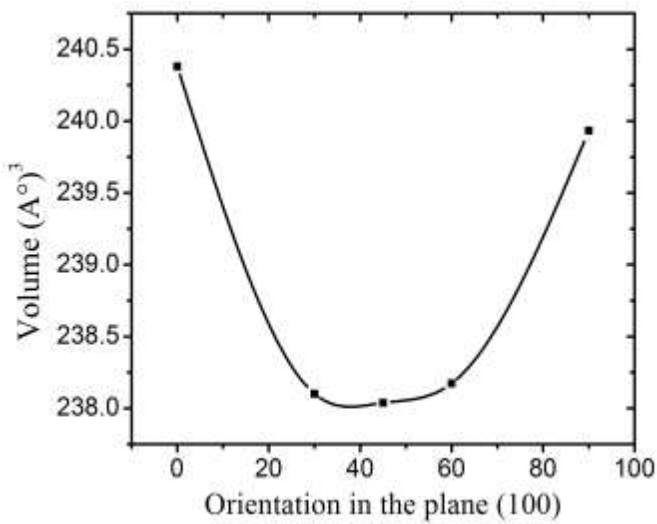

**(b)**

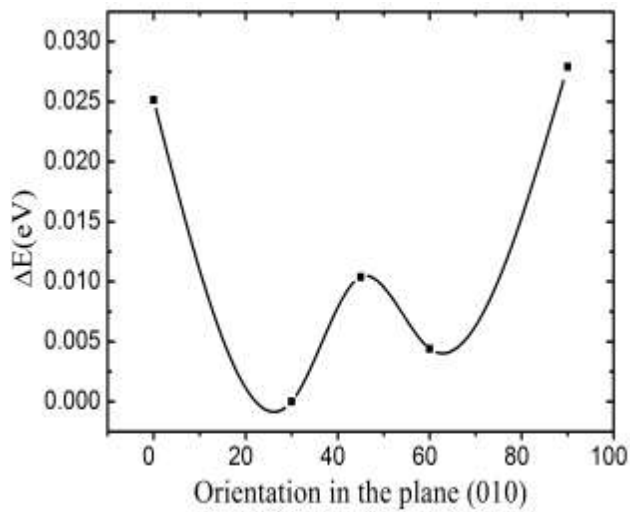
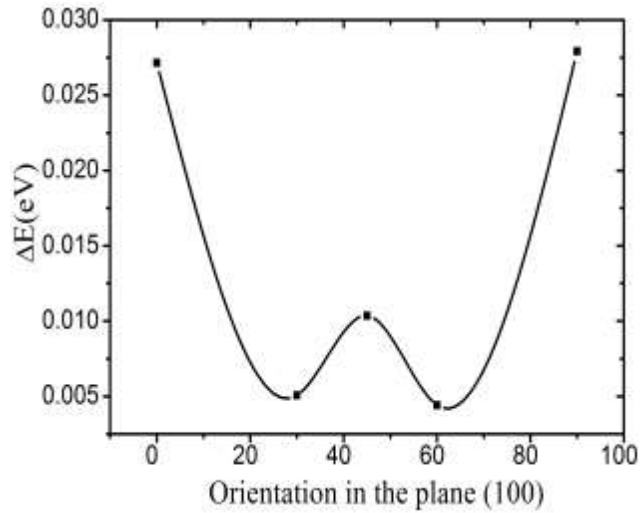

**(c)**

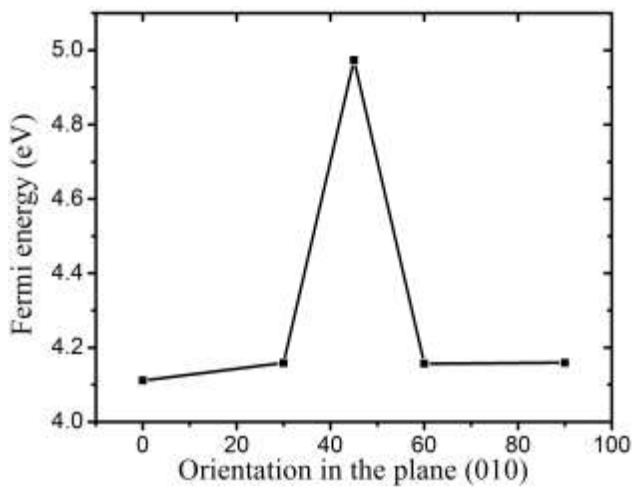
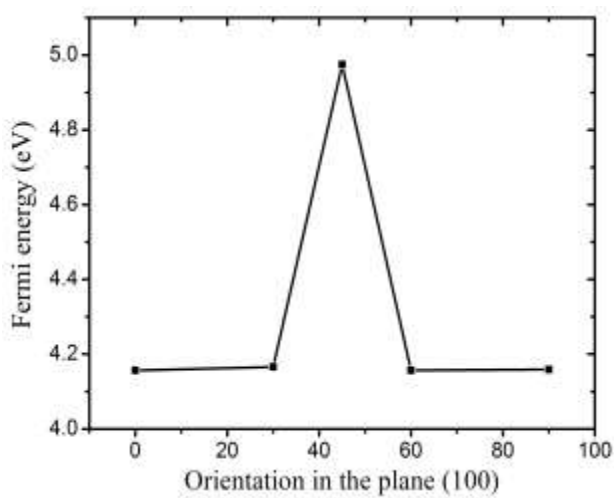

**(d)**

FIG. 5. Values of a) bandgap; b) volume; c) energy barrier; and d) Fermi energy as a function of orientation. Left graphs correspond to the orientation in the plane (010), and the right ones to those in plane (100).

The material MAPbI$_3$ undergoes a first-order phase transition [46,48] from the cubic to the tetragonal phase around 327 K [49]. The thermal energy perturbation at 327 K is $k_BT$=28 meV, which higher than the maximum rotation energy barrier $\Delta E$, As we can see in Table II, which confirms the disordered dynamics of organic cations MA in the cubic phase, consistently with experiments.

### D. Fermi energy versus MA's orientation; open-circuit voltage and ambipolarity

Most photovoltaic cells consist of mainly silicon-based semiconductors. They operate based on the PN diodes, where electrons and holes move in opposite directions, leading to a potential difference across the junction. This potential difference results from the difference in the Fermi energies of the hole-doped and electron-doped semiconductors.

In MAPbI$_3$ hybrid material, as shown in Table II and Fig. 5(d), the Fermi energy $E_F$ depends strongly on the orientation of MA.

The difference between the highest and lowest values of $E_F$, which correspond to orientations 180° wrt [100] and 45° wrt [010], respectively, is about 1 eV. We expect this variation of $E_F$ to induce an electric potential even in the case of an open circuit situation like in the case of a PN junction. In real MAPbI$_3$ materials, molecules MA are not rotating in unison because the correlations between MA dipoles are weak in the cubic phase. This will undoubtedly create a spatial variation in the Fermi energy, causing the material to be ambipolar. An estimate of this potential shall be

$$V_1 - V_2 = \frac{E_F^{MAX} - E_F^{MIN}}{-e},$$

where $e$ is electron's charge, and $E_F^{MAX}$ and $E_F^{MIN}$ are the maximum and minimum values of the Fermi energy, respectively. The voltage calculated in this way, found to be 871 mV, agrees very well with the open-circuit voltage $V_{OC}$ reported experimentally (878 mV) in the tetragonal phase for temperature higher than 300 K [50]. For the cubic phase, our estimate is also in agreement with the experimental value of $V_{OC}$, which is between 0.7 V and 0.8 V [28].

This potential is expected to tear apart electrons and holes of the electron-hole pairs generated in the MA material, hence increasing the lifetime of excitons. This explanation should hold also in the case of MAPbI3 in the tetragonal phase [51]. Our finding is in line with those of Frost *et al*. [9] who proposed that the ferroelectric domains formed by MA molecules create an internal junction which separates the electron and holes [9]. But the fundamental phenomenon responsible for this is not the MA dipole moment but the orientation-induced Fermi level difference. This is in agreement with Chen *et al*.'s earlier statement that the ferroelectric domains are not responsible for the long lifetime, because no domains occur in the cubic phase [52].

According to our results, we can separate all orientations of MA into two categories; the first is the orientation 45° in the two planes **ac** and **bc** (see Table II), and the second is the average of all other orientations (≠45°) because the variation in Fermi energy is negligible as seen in Fig. 5(d). We find that orientation 45° for MA creates an n-type semiconductor, and orientations where molecules MA are aligned to any other direction create a p-type semiconductor. The superposition of these tow semiconductors gives rise to an intrinsic junction.

In the orthorhombic phase, formation of domains with different orientations of MA's electric moments has been reported in literature [53,54]. It is legitimate to assume that these domains have different Fermi energies, and it is thus legitimate to propose that any two adjacent domains behave as a diode with an effective PN junction with two different Fermi energies. Such a junction could be suggested to be at the origin of the nonzero value of the open-circuit potential. We claim that the interface between any two adjacent domains behaves as a junction across which the Fermi energy changes like in a PN junction. This should also provide an explanation for the ambipolarity that was observed in this material since the same sample of material MAPbI3 has parts that behave as p-doped or n-doped [24,25], contrary to material CsPbI3. It was reported that CsPbI3 is a unipolar semiconductor, but the fabrication method or the density of defects may change the nature of the charge carriers [55]. Experimental works reported that it can be an n-type material [56,57], while p-type behavior was predicted theoretically [58,59].

The variation of the open-circuit voltage $V_{OC}$ with temperature has been measured by Patel *et al*. [28]. Practically, $V_{OC}$ does not vary significantly for the wide range of temperatures from 100 to 350 K. For temperatures lower than ~100 K, $V_{OC}$ decreases significantly with decreasing temperature [28]. It continuous to decrease with decreasing temperature for $T < 50$ K, where the material is characterized by long-range ferroelectric order [54]. Below 50 K, in the long-range ferroelectric order regime, one domain is expected to dominate with decreasing thermal fluctuations at the expense of other domains with decreasing temperature. Theoretically, we expect $V_{OC}$ to vanish when only one domain forms in the whole sample. Thus, this is consistent with our explanation for $V_{OC}$ based on the orientation of the MA molecules and ferroelectric domains in the orthorhombic phase.

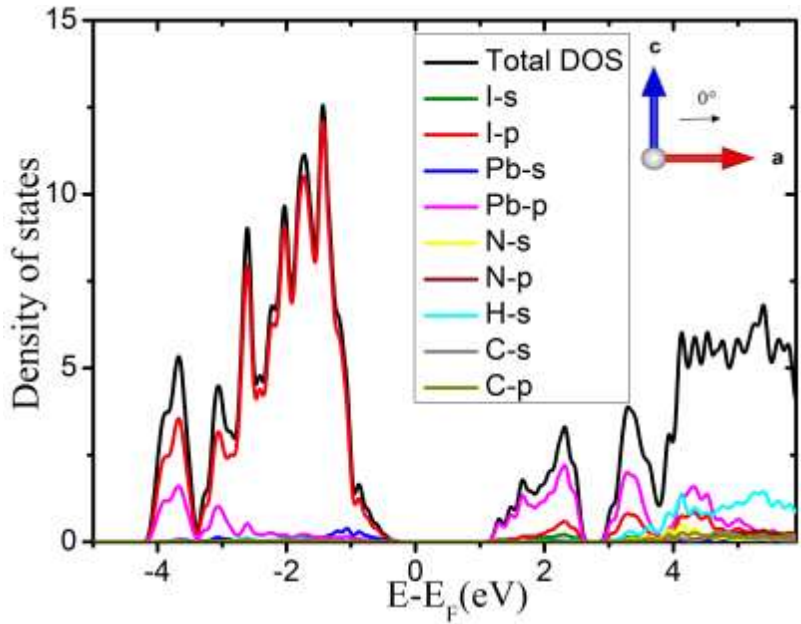

(**a**)

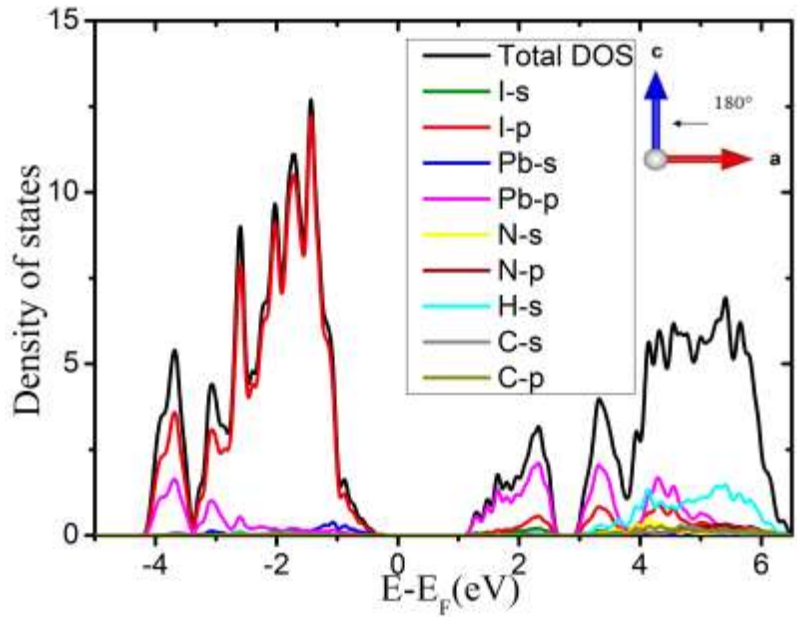

**(b)**

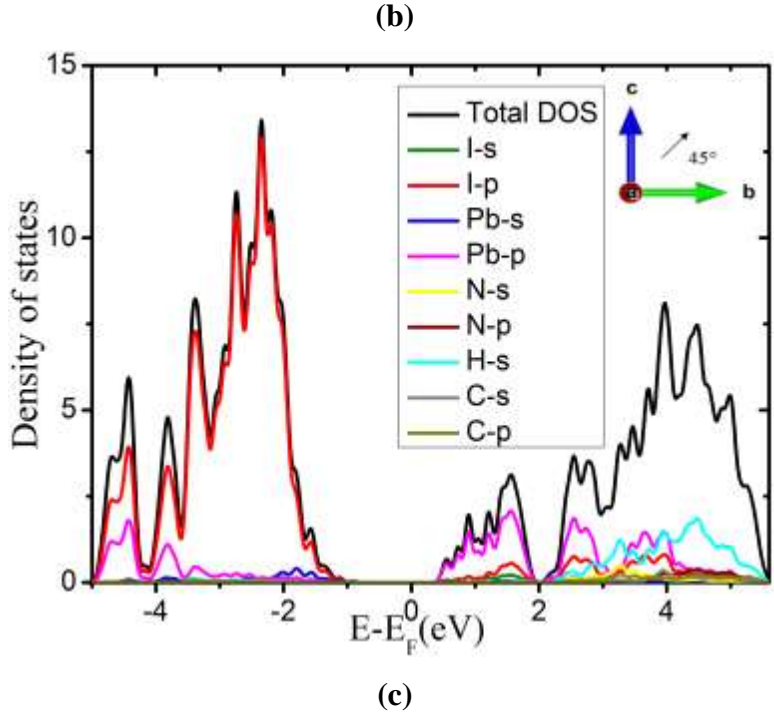

**(c)**

FIG. 6. Total and partial DOS for MA oriented (a) along **a** axis, (b) 180° wrt [100] in plane (010) and (c) 45° wrt [010] in plane (100).

To provide further support to our explanation for the ambipolarity in MAPbI$_3$ [26], Figs. 6(b) and 6(c) display the total and partial DOS for two orientations of MA molecules that show an n-type behavior when the molecule is oriented at 45° with respect to [010] in the plane (100) because the Fermi level is closer to the CBM, and a p-type behavior for MA oriented at 180° with respect to [100] in the plane (010) because the Fermi level is closer to the VBM.

Let us focus on the partial electron densities to show the dominating states around the gap. As we can see from Fig. 6, the band gap boundary states come essentially from the orbitals of Pb and I atoms. The CBM is essentially formed by the p orbitals of Pb. This reflects the ionic nature of hybrid perovskites. The VBM has a strong anti-bonding coupling (Fig. 7) between the s orbital of Pb and p orbital of I. The electronic levels of the organic molecule are located far away from the Fermi level. This agrees with existing literature findings [60].

To show the effect of the organic molecule rotation on the chemical bonding of the MAPbI$_3$ unit cell, we calculated the density difference $\Delta\rho$ (e$^-$/bohr$^3$), which is defined by the difference between the total density $\rho$, which is the sum over all core and valence occupied states of the crystal, and the atomic densities. This difference yields an excellent indication of the orbital overlapping and electron bonding [61].

The organic molecule orbitals do not show up around the gap as we see in Fig. 7, which is consistent with the DOS in Fig. 6. In the valence band, the orbitals of Pb and I are not affected by MA rotation, which is in line with the insignificant variation observed in the valence band structure and on band dispersion width (Table S1). However, in the conduction band, the influence of the MA rotation on the orbitals of the inorganic cage is very important as illustrated by the significant changes observed in the conduction band structure. Besides, the charge density of p orbital of I atom is more intense near the atom N of MA cation because of N high electronegativity and ionic radius [34].

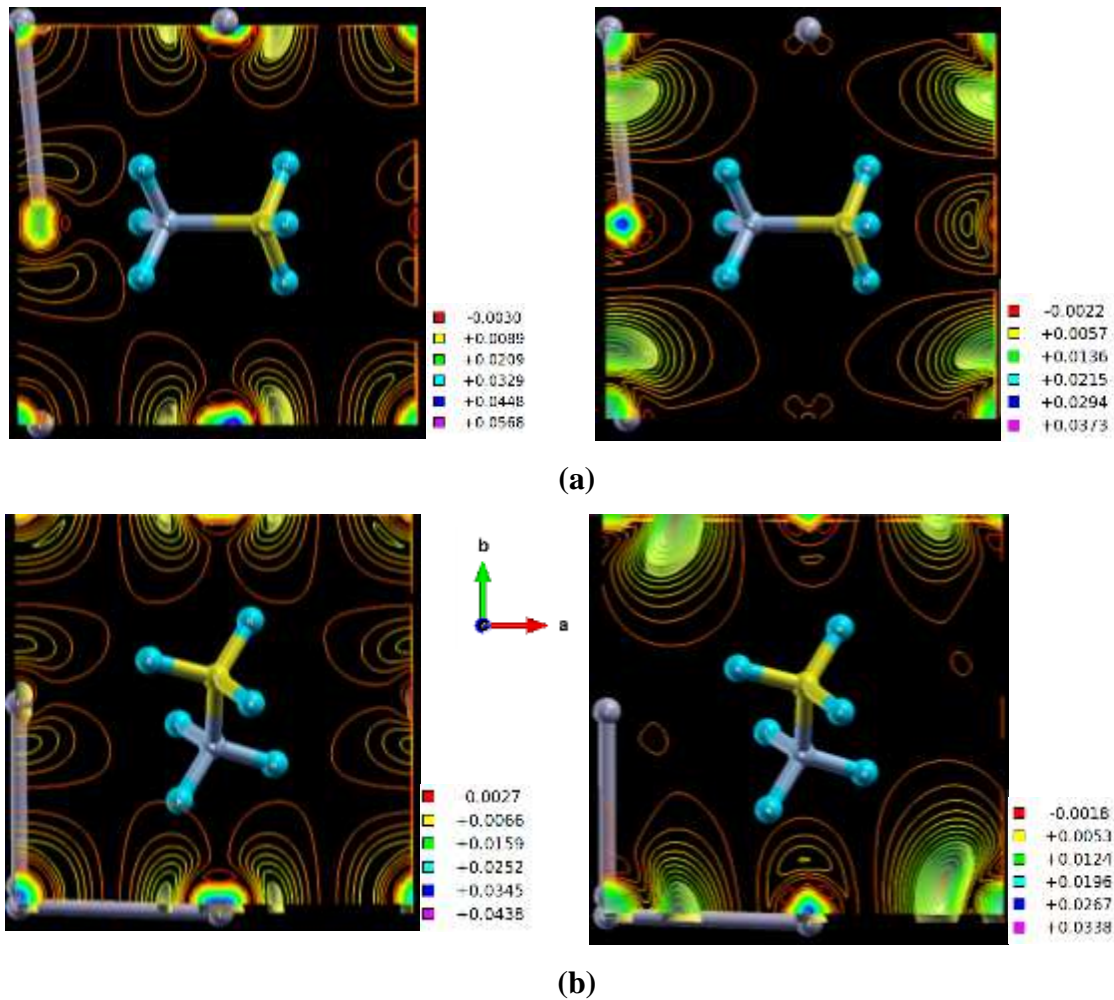

(a)

(b)

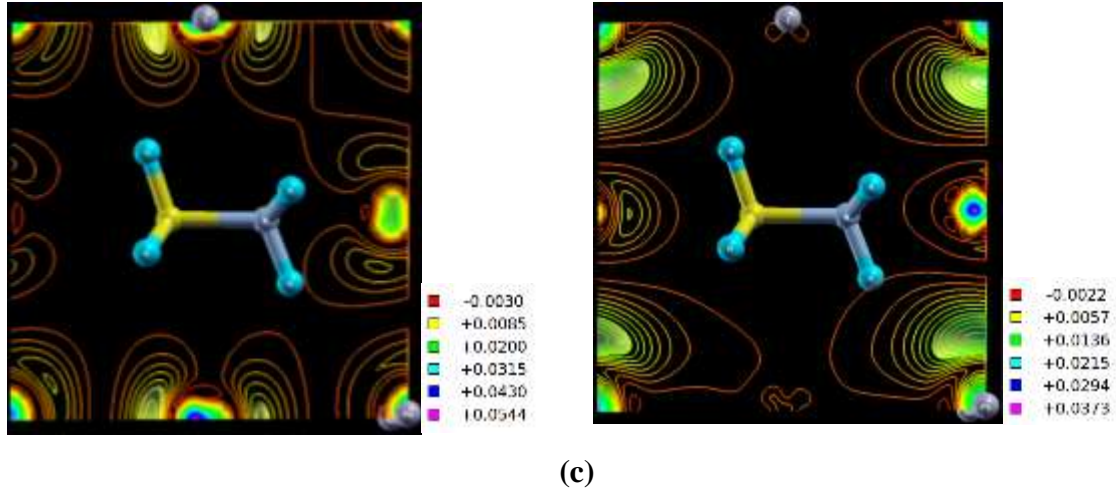

**(c)**

FIG. 7. Density of charge in the Valence Band (left) and Conduction Band (right) at **k**-point R for MA: a) along **a** axis. b) 45° wrt [010] in the plane (100). (c) 180° wrt [100] in the plane (010).

### E. Electric conductivity versus chemical potential

In the relaxation time approximation, Boltzmann's transport equation can be solved at the first order in the external field, and expressions of quantities, such as electrical conductivity can be obtained.

The variation of the chemical potential is equivalent to varying the charge carrier density. The chemical potential has a negative value for holes, and it is defined as $\mu_{hole} = E - E_{VBM}$, where $E < E_{VBM}$. It has a positive value for electrons, $\mu_{electron} = E - E_{CBM}$, where $E > E_{CBM}$ (Fig. 8). We notice that for positive values, i.e. in the conduction band, the electric conductivity increases due to electrons' jet in these bands and assumes a maximum value of $4.116 \times 10^{18}$ $\Omega^{-1}\text{cm}^{-1}\text{s}^{-1}$ at 3.13 eV. For the valence band, the electric conductivity reaches $2.108 \times 10^{18}$ $\Omega^{-1}\text{cm}^{-1}\text{s}^{-1}$ at -2.98 eV. This is due to electrons' extraction from the VB. Our values and features are in good agreement with work reported in Ref. [62].

The diffusion of electrons is much more significant in *xx*, *yy* and *zz* directions than in diagonal directions (*xy*, *xz*, ...). The dominant states around the gap are the orbitals corresponding to Pb and I atoms. These atoms are located on the faces of the unit cell. Thus, electrons move along the principal axes, which explains the greater values of the electric conductivities along these principal directions of the cell.

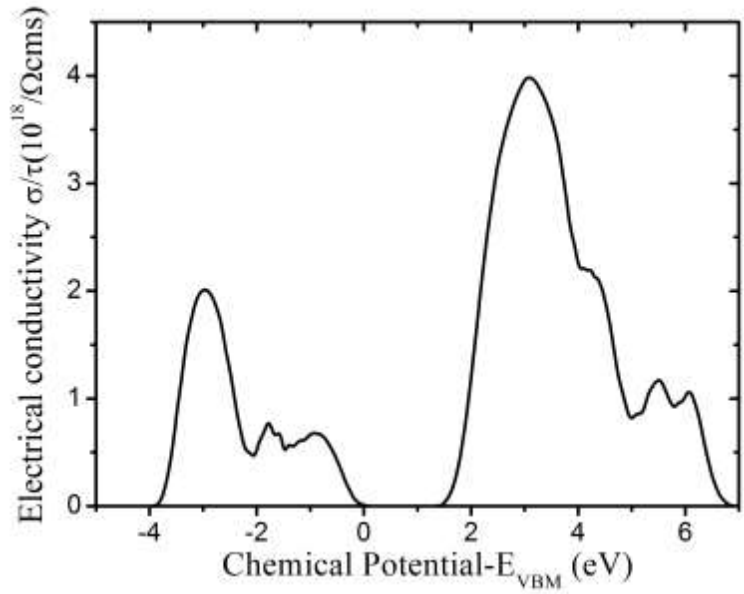

FIG. 8. Electrical conductivity as a function of chemical potential for MA oriented along axis **a**.

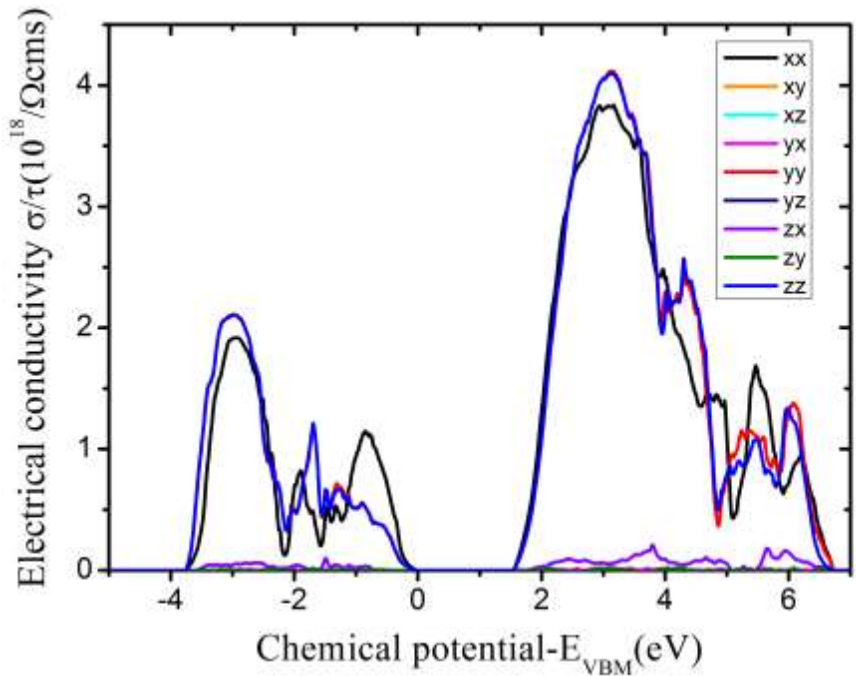

FIG. 9. The variation of electric conductivity for different directions of the cell for MA oriented along **a**.

The conductivity in the cubic phase should be isotropic when averaged on all orientations of MA. In our study, we impose four different directions to observe the effect of the molecule on the conductivity in a unit cell. Fig. 10 displays the electric conductivity at the edge of the valence band. Notice that the electric conductivity is much greater in the direction of the organic cation MA.

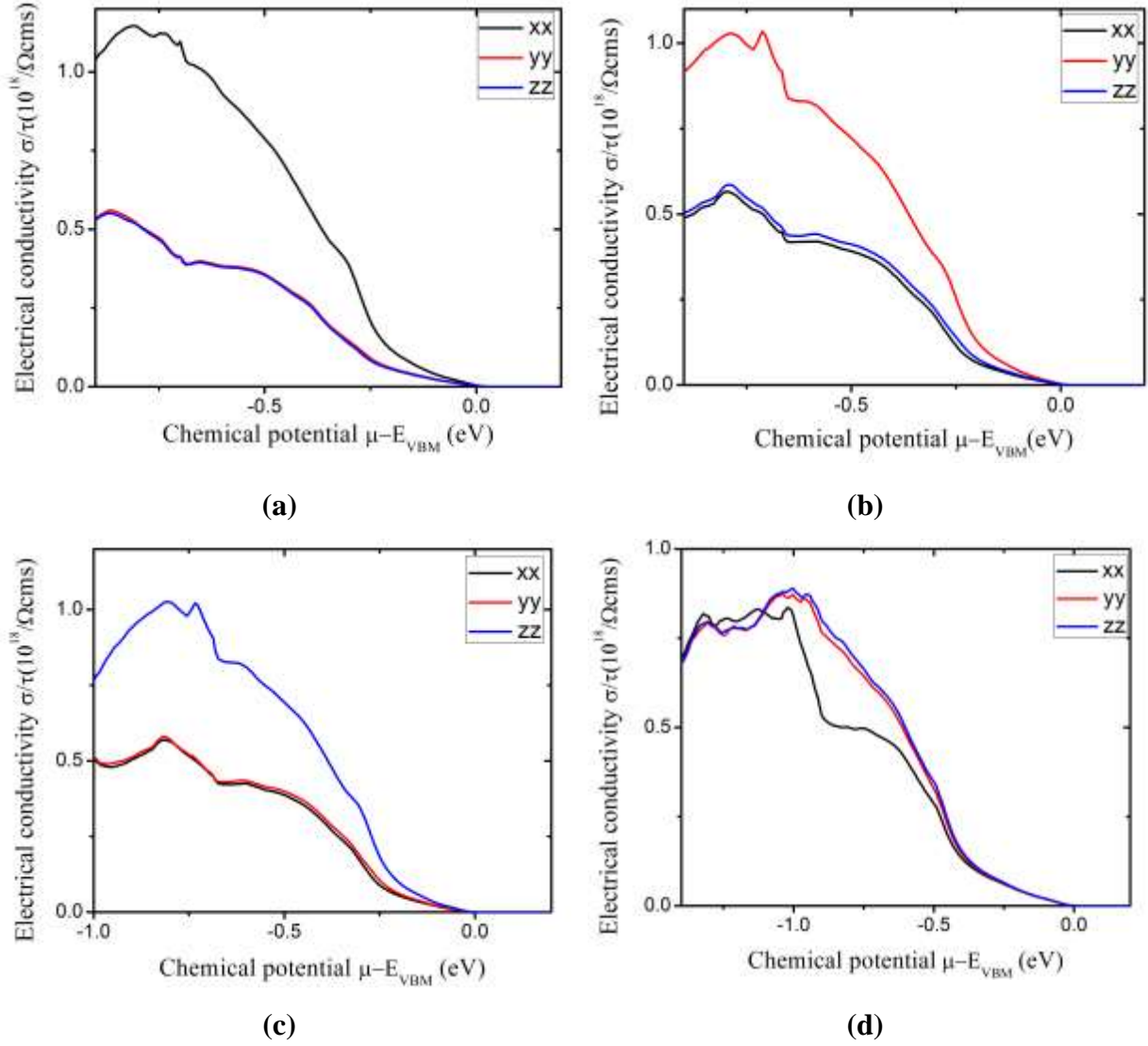

FIG. 10. The variation of the electrical conductivity with MA oriented in the: **a)** [100] orientation; **b)** [010] orientation; **c)** [001] orientation and **d)** [111] orientation.

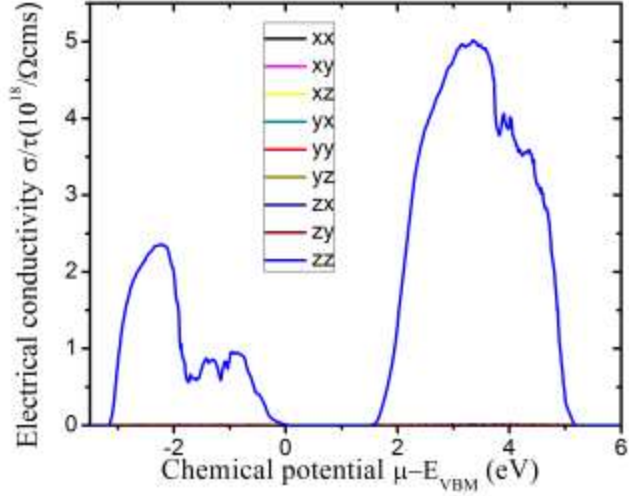

FIG. 11. The variation of electrical conductivity of CsPbI$_3$ as a function of chemical potential for different cell directions.

For comparison purposes and to highlight the important effect of MA, we consider the inorganic perovskite CsPbI$_3$. In this case the electric conductivity is isotropic, i.e. it is the same along the principal directions of the cell. This confirms the important role played by the dipole carried by the organic cation MA in electric conduction and the separation of photo-induced excitons. This explains the important value of electric conductivity along the direction of MA.

## IV. CONCLUSION

In this work, we present strong evidence and arguments that a single mechanism based on the orientations of MA molecules is responsible for the three most important physical properties of hybrid perovskites, i.e., long excitons' lifetime, open-circuit voltage, and ambipolarity. Even though the states related to the organic cation MA in the hybrid perovskite MAPbI$_3$ do not appear near the edges of the bandgap, our results affirm that the orientation of MA influence indirectly all optoelectronic properties of this material. The cell parameters and angles of the crystal structure vary with the rotations of MA, and they thus affect the electronic properties. Most importantly, the rotation-induced Fermi energy variation or simply the dependence of the Fermi energy on the orientation of MA in ferroelectric domains is the key fundamental mechanism that is able to explain the long excitons' lifetime, the origin of the ambipolarity, and the open-circuit voltage. Our calculations show that in the orthorhombic phase, the ferroelectric domains with different average orientations of MA moments should have different Fermi levels, so that any two adjacent domains having different Fermi levels shall act as a PN junction. This explains thus ambipolarity, i.e., a single sample of MAPbI$_3$ may behave simultaneously as a p-type and n-type semiconductor. Also, the difference in the Fermi level is responsible for the open-circuit voltage in the same way as in a conventional diode. This open-circuit voltage is responsible for the long lifetime of photon excited electron-hole pairs (excitons). In the tetragonal and cubic phases where the molecules MA rotate, we speculate that domains or regions of the sample that rotate at different velocities may simulate the same effect of the ferroelectric domains. Another idea is that impurities and/or defects may trap MA molecules at different places in the sample in different orientations. This should have the same effect of two different ferroelectric domains in the orthorhombic phase. These ideas deserve further investigation.